\documentclass[twoside]{article}
\usepackage{fleqn,espcrc2,epsfig}

\usepackage{graphicx}
\usepackage[figuresright]{rotating}

\newcommand{\ba}{\begin{eqnarray}}
\newcommand{\ea}{\end{eqnarray}}
\newcommand{\be}{\begin{equation}}
\newcommand{\ee}{\end{equation}}
\newcommand{\bes}{\begin{equation*}}
\newcommand{\ees}{\end{equation*}}
\newcommand{\bi}{\begin{itemize}}
\newcommand{\ei}{\end{itemize}}

\newcommand{\bcentre}{\begin{center}}
\newcommand{\ecentre}{\end{center}}

\newcommand{\ZV}{Z_V^{\rm eff}}

\font\tenmsb=msbm10 scaled\magstep1
\font\sevenmsb=msbm7 scaled\magstep1
\font\fivemsb=msbm5 scaled\magstep1
\newfam\msbfam
\textfont\msbfam=\tenmsb
\scriptfont\msbfam=\sevenmsb
\scriptscriptfont\msbfam=\fivemsb

\newcommand{\order}[1]{{\mathcal O}(#1)}

\newcommand{\vk}{v\cdot k}

\title{
\hfill\begin{minipage}{0pt}\scriptsize \begin{tabbing}
	\hspace*{\fill} Edinburgh-2000/20\\ \end{tabbing}\end{minipage}\\[8pt]
	\vspace{-1.0cm}
Leptonic and semi-leptonic B decays}

\author{UKQCD Collaboration\\
	Presented by C.M.~Maynard\address{Department of Physics and Astronomy,
	University of Edinburgh EH9 3JZ, UK}} 
\begin{document}

\begin{abstract}
We present results for the semi-leptonic and leptonic decays of $B$
mesons.  These non-perturbative matrix elements are important for
constraining the CKM matrix. Results are presented for the
pseudoscalar and vector decay constants, as well as flavour breaking
ratios and heavy quark symmetry relations. We consider the chiral and
momentum dependence of the semi-leptonic form factors of the decay
$B\to \pi l\nu$ and the soft pion relation (SPR) on the lattice. These
calculations were performed in the quenched approximation at two values
of the coupling with non-perturbatively $\order{a}$ improved
action and currents.
\end{abstract}

\maketitle

\section{INTRODUCTION}
The determination of decay constants and form factors arising from the
decay of B and D mesons is of great importance to phenomenology, in particular
in constraining the Cabibbo-Kobayashi-Maskawa (CKM) matrix. These quantities
can be calculated on the lattice.

The action used in this calculation is the $\order{a}$
non-perturbatively improved action~\cite{alpha_np}, where $c_{SW}$
and the coefficients needed to improve the currents, are determined 
non-perturbatively where possible. The calculation was
done at two values of the coupling, $\beta=6.0$ and $\beta=6.2$.  The
simulation parameters are shown in Table \ref{tab:sim_details}. 
We simulate with three light quark masses around strange and four propagating
heavy quarks around charm. Heavy Quark Symmetry
(HQS) is used to motivate the form of the heavy extrapolation to the
$B$ scale. The light quark mass parameters ($\kappa_{\rm crit}$ etc) have
been determined by the UKQCD collaboration~\cite{QLHS}.
\begin{table}[!ht]
\vspace{-1.0cm}
\begin{center}
\caption{Simulation parameters.}
\label{tab:sim_details}
\begin{tabular}{lcc}
\hline\hline
                            &  $\beta=6.2$        & $\beta=6.0$        \\
\hline
Volume       & $24^{3}\times 48 $& $16^{3}\times 48 $\\   
$c_{\mathrm{SW}}    $       & 1.614       &     1.769                 \\   
$N_{\mathrm{configs}}$      & 216        &     305                  \\
$a^{-1}(r_0)$ GeV & $2.913(10)$  & $2.119(7)$ \\ 
\hline\hline
\end{tabular}
\end{center}
\vspace{-1.1cm}
\end{table}
The vector and axial currents used in this calculation are improved and
renormalised as follows:
\begin{eqnarray}
V_\mu^R&=&Z_V(1+b_Vam_q)\left(V_\mu 
	+ ac_V\tilde{\partial}^\nu T_{\mu\nu}\right) \\
A_\mu^R&=&Z_A(1+b_Aam_q)\left(A_\mu
  + ac_A\tilde{\partial}_\nu P\right) 
\end{eqnarray}

\section{DECAY CONSTANTS}
The results for the decay constants can be found in
\cite{np_imp_fb}, as well as a discussion of the improvement
coefficients, but we present the main results again. The definition of the
vector decay constants differs from the pseudoscalar and so we give it here:
\be
\langle0|V_{\mu}^{\mathrm R }(0)|{\mathrm V }\rangle =
  i \epsilon_{\mu}\frac{M_V^2}{f_V}
\label{eqn:fvdef}
\ee
The decay constants were extracted from appropriate ratios of
correlation functions, and then extrapolated linearly in the light
quark mass to $\kappa_n$ and $\kappa_s$. The extrapolation to the $B$ meson
is motivated by HQS, of the form
\be
\label{eqn:HQS_f}
\Phi_{\rm i}(M_i) \equiv
     C(M_i)f_i\sqrt M_i = 
     \gamma_{\rm i} \left( 1 + \frac{\delta_{\rm i}}{M_i} 
     + \frac{\eta_{\rm i}}{M_i^2} \right)
\ee
where $i$ is $P$ or $V$\footnote{In this case $f_i=M_V/f_V$} 
and the function $C(M_i)$ contains
the leading logarithmic $\alpha_s$ corrections to the scaling
relation

The results for the decay constants are shown in table
\ref{tab:decay_const}.  The large value of the decay constants is in part due 
to the choice of quantity used to set the scale. $f_B$ varies from
$218$ MeV using $r_0$ to set the scale to $186$ using $m_\rho$. This
dominates the systematic uncertainty. The remaining systematic error
comes from difference between $\beta=6.2$ and $\beta=6.0$, and choice
of functional form for the heavy extrapolation (quadratic vs linear).
According to HQS, the $\Phi$ functions for the $P$ and $V$ decay
constants should agree in the heavy quark limit. However, the
$\beta=6.0$ data fail to satisfy this constraint, which may be
indicative of large discretisation errors on the coarser lattice.  To
examine this further we considered, the KLM
norm~\cite{KLM_norm}\footnote{modified to account for $\order{a}$
improvement} and the dispersion relation. At $\beta=6.2$, the KLM
factor was close to unity and the dispersion relation showed good
continuum behaviour. However, at $\beta=6.0$, the KLM factor has a
finite effect, and the dispersion relation shows a deviation from the
continuum version. These all suggest that the $\beta=6.0$ data are
being affected by $\order{(am_Q)^2}$ errors.

\begin{table}[!ht]
\vspace{-0.5cm}
\caption{The decay constants. The first error is statistical, the second
	is systematic.}
\label{tab:decay_const}
\bcentre
\begin{tabular}{ccc}
\hline\hline
& $P$ & $V$\\\hline
$B$& $218(5)^{+\ 5}_{-41}\ {\rm MeV}$&$22.6(0.7)^{+4.4}_{-3.6}$\\
$D$& $220(3)^{+\ 2}_{-24}\ {\rm MeV}$&$\ 7.5(0.1)^{+1.3}_{-0.8}$\\
$B_s$& $242(4)^{+13}_{-48}\ {\rm MeV}$&$20.9(0.4)^{+3.3}_{-4.2}$\\
$D_s$& $241(2)^{+\ 7}_{-30}\ {\rm MeV}$&$\ 7.3(0.1)^{+0.9}_{-0.4}$\\
\hline
$B/B_s$&$1.11(0.01)^{+0.05}_{-0.03}$&$0.92(0.01)^{+0.04}_{-0.03}$\\
$D/D_s$&$1.09(0.01)^{+0.05}_{-0.02}$&$0.98(0.01)^{+0.02}_{-0.04}$\\
\hline\hline
\end{tabular}
\ecentre
\vspace{-1cm}
\end{table}

\section{RENORMALISATION}
\begin{figure}[!ht]
\vspace{-0.5cm}
\bcentre
\epsfig{file=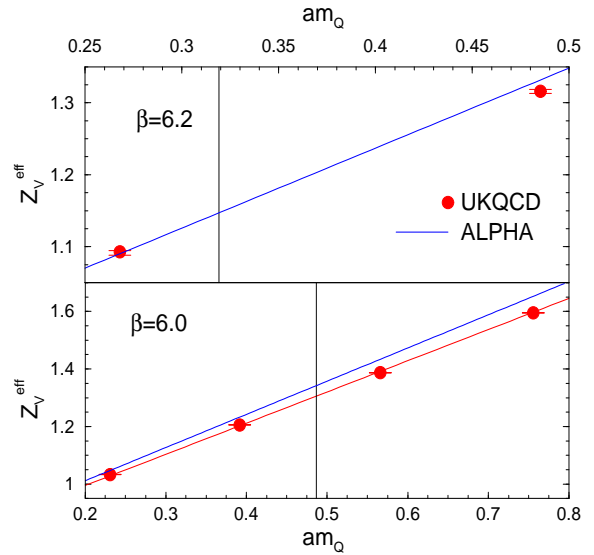,height=7.5cm,width=7.5cm}
\vspace{-1.5cm}
\caption{The mass dependence of $\ZV$. The vertical line shows the charm quark
	mass as determined from the heavy-light pseudoscalar.}
\label{fig:ZV_eff}
\ecentre
\vspace{-1cm}
\end{figure}

The matrix element of a $P\to P$ transition can be parameterised in terms
of two form factors,
\ba
&  Z_V^{\rm eff}\langle P(\vec{p_2})| V_{\mu} | P(\vec{p_1}) 
	\rangle^{\rm latt} &\\\nonumber
	&=f_+(q^2)(p_1+p_2-\Delta_{m^2}q)_\mu + f_0(q^2)\Delta_{M^2}q_\mu &
\ea
where $q=p_1-p_2$ and $\Delta_{m^2}$ is the difference of the square
of the masses normalised by $q^2$.  Recalling that the vector current
is conserved, we know that $f_+(0)$ must be one and therefore consider
the {\em forward, degenerate} matrix element. Here, $\ZV$ can then be
measured directly by taking the appropriate ratio of two- to
three-point correlation functions.  This can then be compared to
$Z_V(1+b_Vamq)$ with the values of $Z_V$ and $b_V$ determined by
\cite{alpha_np3_Z}, at the values of quark mass for which we measure
$\ZV$.  This is shown in figure
\ref{fig:ZV_eff}.

These data have already been shown for $\beta=6.2$ in
\cite{B2pi_plb}. The ALPHA determinations of $Z_V$ and $b_V$ used chiral ward
identities, and so used zero or very small quark mass. Our results
show excellent agreement with the mass-independent scheme even at
large quark mass.

\section{THE SOFT PION RELATION}
HQS can be used to relate the scalar form factor, $f_0$ to the ratio
of the $B$ to pion decay constant,
\be
f_0(q_{\rm max}^2)=\frac{f_B}{f_\pi}
\ee
A recent review~\cite{hashi_lat99} found that the majority of lattice
calculations found this relation violated.  However, the key issue for the
SPR on the lattice is the chiral extrapolation.
This is complicated by the dependence of the form factors on 
$q^2$, itself heavily dependent on the light quark mass. In this calculation
the active light quark (A), which occurs in the heavy-light current and
the spectator light quark (S) which takes no part in the weak decay have
different mass. Hence,
\be
 f=f(q^2_{(m_A,m_S)},m_A,m_S)
\ee
A Taylor expansion of $q^2$ shows $q^2\sim M^{\rm light}_{P_S}$, and
we know from PCAC that  $(M^{\rm light}_{P_S})^2\sim\bar{m}$ where the average
quark mass $\bar{m}=(m_A+m_S)/2$. Thus the light quark mass
dependence of the form factor is
\be
f(q^2,\bar{m},m_S)=\alpha + \beta \bar{m}^{1/2} + \gamma \bar{m} + \delta m_S
\ee
The second term in this expression makes the extrapolation difficult
to control, as it is non-analytic and $\partial \bar{m}^{1/2}/\partial
\bar{m}$ blows up as we approach the chiral limit, in the region where
we have no data.

Instead of extrapolating at fixed pion three-momentum we separate the $q^2$ and
chiral behaviour of the form factor. We
first {\em interpolate} in $q^2$ at fixed quark mass
so that at each quark mass we have the form factors at the same momentum
value, and then extrapolate {\em at fixed $q^2$} without the troublesome
second term.
However, because we extrapolate at fixed $q^2$, we no longer have data
at $q^2_{\rm max}$ for physical quark masses.  This method has already
been successfully employed to calculate the differential decay rate
for $B\to\pi$ at $\beta=6.2$ in~\cite{B2pi_plb}.  In this analysis, an
extra-momentum channel is included ($1\to\ \sqrt{2}_{\parallel}$),
and the range of $q^2$ altered to allow interpolation at fixed quark
mass for both values of $\beta$, in the same range of $q^2$.

\begin{figure}[!ht]
\bcentre
\epsfig{file=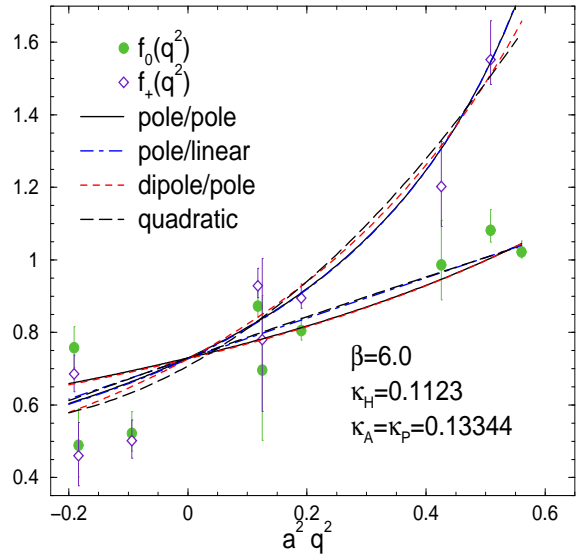,height=7.5cm,width=7.5cm}
\vspace{-1.5cm}
\caption{Interpolation functions for heaviest $\kappa$ combinations at 
	$\beta=6.0$. The heavy quark has a mass around charm.}
\label{fig:inter-func}	
\ecentre
\vspace{-1.0cm}
\end{figure}

The form of interpolating function is motivated by pole-dominance models.
\be
  f_i(q^2)=\frac{f_i}{(1-q^2/M_i^2)}
\ee
with kinematic constraint $f_0(0)=f_+(0)$ imposed. We have also tried other
functions for the interpolation, including dipole/pole for $f_+$ and $f_0$
respectively and with or without the kinematic constraint. This is shown in
figure \ref{fig:inter-func}.
Also shown in figure \ref{fig:inter-func} are other less well
motivated functions, such as linear or quadratic in $q^2$.  As we are
interpolating {\em any} reasonable, smooth function should produce
similar results.

For the extrapolation to the $B$ meson mass, we are again guided by
HQS,
\be
  C(M,M_B)f_i(\vk)M^{n_i/2} = 
	\varepsilon_i\left(1+\frac{\zeta_i}{M}+\frac{\xi_i}{M^2}\right)
\ee
where $n_i={-1,+1}$ when $i={+,0}$ and $\vk$ is the kinematic variable,
\be
  \vk=\frac{M^2 + M_\pi^2 - q^2}{2M}
\ee
The fixed $q^2$ method allows us to choose the $q^2$ values at each heavy
quark mass such that $\vk$ remains constant during the heavy quark 
extrapolation.

\begin{figure}[!ht]
\bcentre
\epsfig{file=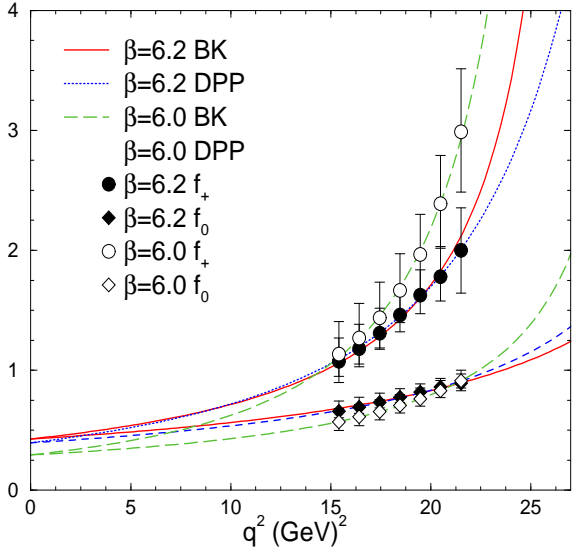,height=7.5cm,width=7.5cm}
\vspace{-1.5cm}
\caption{The momentum dependence of the form factors for $B\to\pi$.}
\label{fig:B2pi_q2}	
\ecentre
\vspace{-1.0cm}
\end{figure}

We can now consider the momentum dependence of the form factors at the
$B$ scale. Combining HQS scaling relations and pole dominance models
suggest the momentum dependence is a dipole for $f_+$ and a pole for
$f_0$ (DPP). A more physical model has been suggested by Becirevic and
Kaidalov (BK)~\cite{BK_param}
\begin{eqnarray}
  f_+(q^2)&=&\frac{c_B(1-\alpha)}
  {(1-q^2/M^2_{B^{\star}})(1-\alpha q^2/M^2_{B^{\star}})} \nonumber \\
  f_0(q^2)&=&\frac{c_B(1-\alpha)}{(1- q^2/\beta M^2_{B^{\star}})}.      
\end{eqnarray}
These are shown in figure \ref{fig:B2pi_q2}.
The value obtained for $f_0(q^2_{\rm max})$ can then be compared to
the value of $f_B/f_\pi$. To try and estimate the systematic
errors we considered the following; $r_0$ versus $m_\rho$ to set the
scale, $\beta=6.2$ versus $\beta=6.0$, different models for the fixed 
$q^2$ interpolation, quadratic vs linear heavy extrapolation. We see the
SPR satisfied on the lattice, with large systematic errors,
but with a tendency for $f_0(q^2_{\rm max})$ to lie below $f_B/f_\pi$.

\begin{figure}[!ht]
\bcentre
\epsfig{file=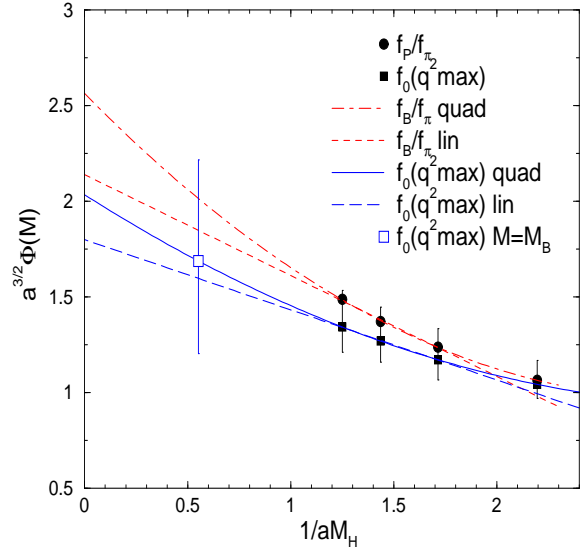,height=7.5cm,width=7.5cm}
\vspace{-1.5cm}
\caption{The soft pion relation}
\label{fig:SPR}	
\ecentre
\vspace{-1.0cm}
\end{figure}

We can also check to see if the extrapolations in $q^2$ and heavy quark mass
commute.  In figure \ref{fig:SPR} the heavy extrapolation of $f_B/f_\pi$ and
$f_0(q^2_{\rm max})$ are shown after $f_0(q^2)$ has been extrapolated with
a pole model to $q^2_{\rm max}$.
Again we see the SPR is satisfied. 

A recent calculation \cite{Onogi_2000} using a NRQCD formulation
of heavy quarks confirms the difficulties with the chiral
extrapolations of the form factors and also adopts a fixed kinetic
variable approach, which reduces their observed violation of the SPR. Another
calculation \cite{Becirevic_2000} using the relativistic heavy Wilson
formulation, and a fixed kinematic variable approach see the SPR satisfied 
if a quadratic heavy extrapolation is used.

The SPR is rather difficult to satisfy on the lattice because we have to
extrapolate the light quark mass into a region where $q^2$ changes rapidly.
This can be circumvented by using the fixed $q^2$ approach. However, 
lattice calculations can reliably determine the differential decay rate for
$B\to\pi$ away from $q^2_{\rm max}$ where the phase space is non-zero to 
extract $V_{\rm ub}$. This work was supported by EPSRC grant GR/K41663 and
PPARC grants PPA/G/S/1999/00022 and PPA/GS/1997/00655.
CMM acknowleges PPARC grant PPA/P/S/1998/00255.

\end{document}